\documentclass[prl,a4paper,nofootinbib,
twocolumn,
]{revtex4-1}
\pdfoutput=1
\usepackage{graphicx}   
\usepackage{bbold}  
\usepackage{mathtools}
\usepackage{amsmath}
\usepackage{amsfonts} 
\usepackage{amssymb}
\usepackage{amsbsy} 
\usepackage{amsfonts}
\usepackage{physics} 
\usepackage{bbold}
\usepackage{slashed}  
\usepackage{color}
\usepackage{tabularx}
\usepackage{hyperref} 
\usepackage{orcidlink}
\usepackage{bbding} 
\usepackage{ulem}

\definecolor{oucrimsonred}{rgb}{0.6, 0.0, 0.0} 
\definecolor{DarkGray}{gray}{0.4}
\definecolor{forestgreen}{rgb}{0.13,0.35,0.13}
\definecolor{ocre}{HTML}{F16723}

\usepackage{bm}
\usepackage{mathrsfs}

\def\eq#1{{Eq.~(\ref{#1})}}
\def\eqs#1#2{{Eqs.~(\ref{#1})--(\ref{#2})}}

\def\abs#1{\left| #1\right|}

\def\Tr{\mbox{Tr}\,}

\colorlet{grayline}{gray!70}
\definecolor{blueline}{rgb}{0,0.27,0.55}
\definecolor{DarkGray}{gray}{0.4}
\definecolor{Gray}{gray}{0.6}
\definecolor{oucrimsonred}{rgb}{0.6, 0.0, 0.0}
\definecolor{persianblue}{rgb}{0.11, 0.22, 0.73}
\definecolor{forestgreen}{rgb}{0.13,0.35,0.13}
 \hypersetup{colorlinks, citecolor=forestgreen, linkcolor=forestgreen, urlcolor=forestgreen}
\newcommand{\be}{\begin{equation}}
\newcommand{\ee}{\end{equation}}
\newcommand{\bea}{\begin{eqnarray}}
\newcommand{\eea}{\end{eqnarray}}
\newcommand{\nn}{\nonumber}

\newcommand*\xbar[1]{%
  \hbox{\;%
    \vbox{%
      \hrule height 0.5pt 
      \kern0.5ex
      \hbox{%
        \kern-0.25em
        \ensuremath{#1}%
        \kern-0.07em
      }%
    }%
  }%
} 
\newcommand{\com}[1]{}
\newcommand{\gsim}{\lower.7ex\hbox{$\;\stackrel{\textstyle>}{\sim}\;$}}
\newcommand{\lsim}{\lower.7ex\hbox{$\;\stackrel{\textstyle<}{\sim}\;$}} 

\newcommand{\bc}{\begin{center}}
\newcommand{\ec}{\end{center}}

\begin{document}

\hypersetup{citecolor = forestgreen,
linktoc = section, 
linkcolor = forestgreen, 
urlcolor = forestgreen
}

\title[]{ \Large \color{oucrimsonred} \textbf{ 
Quantum pions
} }

\author{\bf M. Fabbrichesi$^{a\, \orcidlink{0000-0003-1937-3854}}$}
\author{\bf   R. Floreanini$^{a}\, \orcidlink{0000-0002-0424-2707}$}
\author{\bf E. Gabrielli$^{{b,a,c\, \orcidlink{0000-0002-0637-5124}}}$} 
\author{\bf L. Marzola$^{{c,d\, \orcidlink{0000-0003-2045-1100}}}$}
\affiliation{$^{a}$INFN, Sezione di Trieste, Via Valerio 2, I-34127 Trieste, Italy}
\affiliation{$^{b}$Physics Department, University of Trieste, Strada Costiera 11, I-34151 Trieste, Italy}
\affiliation{$^{c}$Laboratory of High-Energy and Computational Physics, NICPB, R\"avala 10,  10143 Tallinn, Estonia}
 \affiliation{$^{d}$Institute of Computer Science, University of Tartu, Narva mnt 18, 51009 Tartu, Estonia}

\date{\today}

\begin{abstract}
\noindent	We show that two- and three-pion states produced in the decay of neutral kaons are contextual, entangled, and Bell nonlocal in isospin space. By reinterpreting the experimental values of the different isospin amplitudes, we can determine the amount of entanglement enjoyed by these states and the extent to which they violate the non-contextuality and Bell locality inequalities. Notably, the three-pion state offers a genuine multipartite test of Bell nonlocality with qutrits. 
 
\end{abstract}

\maketitle

\textbf{Introduction---} Particles produced in scattering and decay processes can become entangled in their spin state. Similarly, internal space degrees of freedom, such as isospin in flavor space, can also become entangled.

Pions carry isospin $I=1$ and the three pion states, $\pi^+$, $\pi^-$ and $\pi^0$, are members of an isospin triplet. In analogy with the case of the spin of a massive spin-1 particle, we can discuss the quantum properties of the pion isospin triplet. We expect the two- and three-pion final states from the CP conserving decays of neutral kaons $K_S$ and $K_L$ to show quantum contextuality~\cite{Bell:1964fg,Kochen:1968zz}, entanglement~\cite{Horodecki:2009zz} and Bell nonlocality~\cite{Bell:1964}---and the amount of these properties to depend on the relative weight between the different isospin channels regulating the decays. These properties show that pions are proper quantum entities that cannot be visualized, let alone described, as classical point particles.

Once CP-violating effects are taken into account, the actual final states are incoherent mixtures containing a small component of a three-pion state in the predominantly two-pion state produced by the $K_S$ decay, and vice versa for the $K_L$. However, the weight of these extra components is of the order of $10^{-3}$ in both cases, so they can be safely ignored for our purposes.

For the quantum properties to manifest, it is crucial to model the isospin using a non-Abelian group.  Properties of particles that are defined by internal spaces described by Abelian groups---like the electric charge---do not display quantum correlations: they are described by one-dimensional representations which do not allow  different choices of orientations in their internal space. For this reason, for instance, the state of two opposite charged particles (otherwise identical) is described by a factorizable state and the correlations between the two charge states remain purely classical.

The quantum properties we discuss characterize the states of the two- and three-pion systems. These states can be reconstructed by means of quantum state tomography---which is a particularly simple task to achieve in this case.  We utilize the experimental values of the different isospin amplitudes that regulate  the decays to compute the amount of entanglement, contextuality and Bell nonlocality in isospin space.  The three-pion state is of particular interest because it provides a multipartite test of Bell nonlocality with qutrits.

While these quantum properties of the states are indeed real and influence the physical processes, it is crucial to recognize  that there is a difference with respect to when the same quantities are measured in spin space.  For instance, the various average values entering a
Bell inequality cannot be directly measured experimentally because we cannot rotate our apparatus in isospin space, contrary to what we do with polarimeters. Instead, as discussed below, Bell tests can be performed using quantum state tomography in isospin space~\cite{Fabbrichesi:2025aqp}.

To the best of our knowledge, this is the first experimental measure of entanglement in isospin space. Correlations due to Bose symmetry in pion scattering were analyzed in~\cite{Goldhaber:1960sf} (for a review with more recent results see~\cite{Boal:1990yh,Alexander:2003ug}). Such correlations differ from those considered here because they are generated by an incoherent superposition of states. More recently,
entanglement in flavor space has been discussed in scattering processes in~\cite{Beane:2018oxh,Beane:2021zvo,Low:2021ufv,Liu:2022grf,Carena:2023vjc,Kowalska:2024kbs} to explore possible connections between entanglement and   symmetry enhancements.

\vskip0.2cm
\textbf{Two-pion state---} The two pions emerging from the decay of a neutral kaon $K^0$ in its rest reference frame are described by the state
\be
\ket{\Psi} =  \frac{1}{\sqrt{2+|\gamma|^2}}  \qty[\ket{\pi^+\pi^-} + \gamma \ket{\pi^0\pi^0} + \ket{\pi^- \pi^+}]\,,
\label{2state}
\ee
in which the two pions are distinguished by their opposite momenta. The configuration of the state is determined by the requirement $Q=I_z=0$ of the final system as a whole. Neglecting $CP$ violating effects, the two states $ \ket{\pi^+\pi^-}$ and $ \ket{\pi^-\pi^+}$ enter with the same weight and therefore one constant, $\gamma$, is sufficient to fully determine the state. The state in \eq{2state} is formally the same as that describing the final spin state $J=J_z=0$ in the decay of a scalar state into two massive spin-1 states: the $\pi^+$ and $\pi^-$ play respectively the role of the $+1$ and $-1$ components of spin as measured along an arbitrary direction, whereas the $\pi^0$ plays that of the component 0.

Why is the state in \eq{2state} a coherent superposition rather than a mixture of the  $\pi^0\pi^0$ and $\pi^+\pi^-$ pairs? The pions are produced at the leading order by the weak current-current operator that transforms the $s\bar d$ quarks of the kaon into  a $u$ and  $\bar u$ quarks, which are then hadronized by either a pair of $\bar u u$ or $\bar d d$ quarks. The two possibility are to be coherently summed and the $\pi^0\pi^0$ and $\pi^+\pi^-$ final state produced as a superposition.  

The density matrix $\rho=\dyad{\Psi}$ obtained from \eq{2state}
satisfies $\rho^{2}=\rho$, it being a pure state.
The reduced density matrix of one pion, obtained by tracing over the degrees of freedom of the other, is given, 
in both cases, by
\be
\rho_1=\frac{1}{| \gamma | ^2+2} \left(
\begin{array}{ccc}
1& 0 & 0 \\
 0 &| \gamma | ^2& 0 \\
 0 & 0 &1 \\
\end{array}
\right)\, ,\label{red2}
\ee
and describes an incoherent mixture of the three physical $\pi^+$, $\pi^0$ and $\pi^-$ states with weights given by the elements on the diagonal.

The reduced matrix in \eq{red2} determines whether the single pions produced in the decay features quantum contextuality or not by means of  the non-contextuality  inequality~\cite{Fabbrichesi:2025rsg}
\be
\overline{\mathbb{CNTXT}}_9 = \sum_{k=1}^{4} \lambda^{\rho_1}_k \, \lambda^\Pi_k\,  \leq 3\,, \label{eq:max}
\ee
in which $\lambda_i^{\rho_1}$ are the eigenvalues of the matrix $\rho_1$  and $\lambda_i^\Pi$ those of the matrix 
\be
\Pi=\begin{pmatrix} 10/3&0&0\\ 0&3 & 0\\ 0&0& 8/3\end{pmatrix}\, ,
\ee 
combined as to obtain the largest value in the sum in \eq{eq:max}.

We find that
\be
 \mathbb{CNTXT}_9 =
 \begin{cases}
 \dfrac{8 \abs{\gamma}^2+19}{3 \abs{\gamma}^2+6}\text{ if }\abs{\gamma}^2 <1, \\
 \\
 \dfrac{10 \abs{\gamma}^2+17}{3 \abs{\gamma}^2+6} \text{ if }\abs{\gamma}^2 \geq 1,
\end{cases} \label{cont9two}
 \ee
which can be used to determine whether the pions are in a non-contextual state or not.

As the pions are in a pure state, entanglement can be quantified via the entanglement entropy~\cite{Horodecki:2009zz}, equal to
\begin{multline}
\mathscr{E}[\rho]= \frac{1}{| \gamma |
   ^2+2} \Big[ | \gamma | ^2 \log \left(\frac{| \gamma | ^2+2}{|\gamma|^2}\right)  \\
   + 2  \log \left(| \gamma | ^2+2\right) \Big]\, ,
\end{multline}
which can vary from 0 to $\log 3$. The quantity
\be
{\cal I}_3=\Tr[ \mathscr{B} \rho] \, ,
\ee
with the matrix $\mathscr{B}$ given by~\cite{Acin:2002zz}
\be\scriptsize
\mathscr{B} =  \begin{pmatrix} 
  0 & 0 & 0 & 0 & 0 & 0 & 0 & 0 & 0  \\
  0 & 0 & 0 & -\frac{2}{\sqrt{3}} & 0 & 0& 0 & 0 & 0  \\
  0 & 0 &0 & 0 & -\frac{2}{\sqrt{3}} & 0 &2 & 0 & 0  \\
  0 &  -\frac{2}{\sqrt{3}} & 0 & 0 & 0 & 0 & 0 &0 & 0  \\
  0& 0 & -\frac{2}{\sqrt{3}} & 0 & 0 & 0 & -\frac{2}{\sqrt{3}} & 0 &0  \\
  0 & 0 & 0 & 0 & 0 & 0 & 0 &  -\frac{2}{\sqrt{3}} & 0  \\
  0 & 0 & 2 & 0 & -\frac{2}{\sqrt{3}} & 0 &0& 0 & 0  \\
  0 & 0 & 0 & 0 & 0 &  -\frac{2}{\sqrt{3}} & 0 & 0 & 0  \\
  0 & 0 & 0 & 0 & 0& 0 & 0 & 0 & 0  
\end{pmatrix} \,,
\ee
can be used to test Bell locality for two qutrits by means of the inequality
 \be
 {\cal I}_3\leq2\, . \label{bell2}
 \ee
The maximum attainable value within Quantum Mechanics (QM) is $ {\cal I}_3\simeq2.915$~\cite{Acin:2002zz}, whereas in completely nonlocal theories it is $ {\cal I}_3=4$~\cite{Collins:2002sun}.

To determine the value of the entanglement entropy and of the two inequalities in \eq{cont9two} and \eq{bell2}, we must determine the value of the coefficient $\gamma$ entering the density matrix $\rho$.  Assuming CP conservation, the kaon decay process is regulated by the real isospin amplitudes $A_0$ and $A_2$, which are related to the decay modes as
\bea
\langle \pi^+\pi^-|K^0\rangle &=& A_0 \,e^{i \delta_0} + \frac{A_2}{\sqrt{2}} \,e^{i \delta_2}\nn \\
\langle \pi^0\pi^0|K^0\rangle &=& A_0 \,e^{i \delta_0} - \sqrt{2} A_2\,  e^{i \delta_2}\,,\label{amps}
\eea
and determine the value of $\gamma$ as
\be
\gamma = \frac{\langle \pi^0\pi^0|K^0\rangle }{\langle \pi^+\pi^-|K^0\rangle} = \frac{ \sqrt{2}A_0 \,e^{i (\delta_0- \delta_2)} -2 A_2}{\sqrt{2}A_0  \,e^{i (\delta_0- \delta_2)}+A_2} \, .
\ee
If the $I=2$ amplitude is neglected, $\gamma=1$ and the two pions are in the $I=0,\, I_{z}=0$ eigenstate that maximizes the isospin entanglement.
The introduction of an $I=2$, $I_{z}=0$ component, suppressed by the $\Delta I=1/2$ rule, causes a small departure from the $\gamma=1$ limit and enhances the nonlocal character of the state. Had the $I=2$ component being dominant, the state would have been less entangled. The phases in \eq{amps} come from the strong interactions of the final state pions which undergo $\pi$-$\pi$ scattering. It is important to stress that both electroweak (causing the kaon decay) and strong (in the final state scattering) interactions are responsible for the presence of entanglement in the pion final state. For this reason, QM is here been tested in the presence of the full Standard Model.

By using the isospin amplitudes  given in~\cite{Cirigliano:2011ny,FlaviaNetWorkingGrouponKaonDecays:2010lot} 
\bea
A_0 &=& (2.704 \pm 0.001) \times 10^{-7} \; \text{GeV}\nn \\
A_2 &=& (1.210 \pm 0.002) \times  10^{-8} \; \text{GeV}\nn \\
\delta_0 (m_K) &-& \delta_2(m_K) = (47.5\pm 0.9)^\circ \, ,
\eea
we obtain
 \be
\abs{\gamma}^2=0.880\pm 0.002\,. 
\label{gamma} 
\ee
The quantum properties depend strongly on the coefficient $\gamma$ and for large values of the parameter the system becomes disentangled because the components containing charged pions are suppressed. In the opposite limit, $\gamma\to0$, entanglement remains and involves the charged pions only. For the value in \eq{gamma} we find
\be
\mathscr{E}[\rho]=1.097 \pm 0.001\,,   
\ee
close to the largest value achievable, $\log 3 \simeq 1.099$ (obtained for $\abs{\gamma}^2=1$).

Taking the same value for $\gamma$, we find that
 \begin{align}
 \mathbb{CNTXT}_9&=3.014\pm 0.001\,,
 \\
 {\cal I}_3 & =2.890 \pm 0.001 \, .
 \end{align}
The non-contextuality and Bell locality inequalities in \eq{cont9two} and  \eq{bell2} are both violated with a significance well in excess of the $5\sigma$ level. The state in \eq{2state} is thus certified to be non-locally entangled and contextual. The large significances obtained in all these tests are driven by the precision achieved in the experimental determination of the CP conserving kaon decay amplitudes.  In addition, as a value larger than  ${\cal I}_3=2.915$ is not reached, QM is confirmed. 
 
If isospin breaking corrections are included, the values of the isospin amplitudes change~\cite{FlaviaNetWorkingGrouponKaonDecays:2010lot},
but not sufficiently to have an appreciable  impact on the quantum properties of the two-pion state.

\vskip0.2cm
\textbf{Three-pion state---} The three pions coming from the $K_{L}$ decay are describe by the state 
\begin{align}
&\ket\Phi = \frac{1}{\sqrt{6+\abs{\Gamma}^2}} \Big[ \ket{\pi^0\pi^+\pi^-} + \ket{\pi^+\pi^0\pi^-} + \ket{\pi^+\pi^-\pi^0}\nonumber\\ &+\Gamma\,  \ket{\pi^0\pi^0\pi^0} + \ket{\pi^0\pi^-\pi^+}  +\ket{\pi^-\pi^0\pi^+} + \ket{\pi^-\pi^+\pi^0} \Big]\,, 
\label{3state}
\end{align}
in which the pions are distinguished by their momenta $p_{i}$, with $\sum_{i }p_{i}=p_{K}$ (with $p_{K}$ being the kaon momentum).
The relative weight $\Gamma$ of the $\ket{\pi^0\pi^0\pi^0}$ state is determined by the experimental values of the amplitudes $A(\pi^0\pi^0\pi^0)$ and  $A(\pi^+\pi^-\pi^0)$; it depends on the two Dalitz plot variables describing the process.  
The corresponding density matrix $\eta = \dyad{\Phi}$ satisfies $\eta^2=\eta$, it being a pure state. 

Contextuality can be tested for each pion produced in the decay by considering the corresponding reduced density matrices
\be
\eta_0=\eta_{-}=\eta_{+} = \frac{1}{| \Gamma | ^2+6} \left( 
\begin{array}{ccc}
 2 & 0 &0 \\
 0 & | \Gamma | ^2+2 & 0 \\
0& 0 &2 
\end{array}
\right)\,,
\ee
obtained from $\eta$ by tracing over the degrees of freedom of the remaining two pions. 

The non-contextuality inequality, as given by \eq{eq:max}, is found to be
\be
 \mathbb{CNTXT}_9 =
 \dfrac{ 10\, |\Gamma|^2 + 54}{3\, (6 + |\Gamma|^2)} \, ,\label{cont9three}
 \ee
and it is the same for each of the pions.

The classification of entanglement in multipartite systems is richer than in the bipartite case (see, for instance, 
\cite{Horodecki:2009zz,Coffman:1999jd,PhysRevA.83.062325}), and a concept of particular importance is that of genuine multipartite entanglement (GME):
a state features GME when it cannot be decomposed as a convex combination of bi-separable states 
(states which are separable on, at least, one bipartition of the system), and thus exhibits quantum correlations involving
all its parts \cite{PhysRevD.35.3066}. In the case of the three-pion state (\ref{3state}), GME can be quantified by
\be
\mathscr{C}[\eta]=\min \big\{ \mathscr{C}_{1(23)},\, \mathscr{C}_{2(13)}\, \mathscr{C}_{3(12)} \big\}\ ,
\ee
where
\be
\mathscr{C}_{i(jk)}=\sqrt{2 \qty[1-\Tr(\eta_{(jk)}^2)]}\, ,
\ee
in which $\eta_{(jk)} = \Tr_{\!(jk)} [\eta$] is the reduced density  matrix obtained by tracing out two of the pions.
We find
\be
\mathscr{C}_{1(23)} =\mathscr{C}_{3(12)}=\mathscr{C}_{2(31)}=\frac{4 \sqrt{| \Gamma | ^2+3}}{| \Gamma | ^2+6}\, ,
\ee
which, in the present case, quantify the entanglement of one of the pion with the system composed by the other two.

Another quantitative measure of tripartite entanglement  is provided by the concurrence triangle~\cite{PhysRevLett.127.040403,Jin:2022kxb}, given by
\be
F_3=4 \sqrt{\frac{ Q \big(Q-\mathscr{C}_{1(23)} \big) \big(Q-\mathscr{C}_{2(13)} \big) \big(Q-\mathscr{C}_{3(12)}\big)}{3}} \, ,
\ee
in which
\be
Q=6 \frac{ \sqrt{| \Gamma | ^2+3}}{| \Gamma | ^2+6}\,,
\ee
yielding
\be
F_3 = 16 \frac{\abs{\Gamma}^2+3}{(\abs{\Gamma}^2+6)^2}
\ee
in the case of the three pions.
As in the case of two pions, entanglement among the three pions  only vanishes for $\abs{\Gamma}\to\infty$.

Genuine multipartite Bell nonlocality can be probed in a system composed by three qutrits by means of an  operator $\mathscr{B}'$ defined as~\cite{PhysRevD.35.3066,PhysRevLett.89.060401}
\begin{multline}
 \mathscr{B}' =  A_0 \otimes B_0 \otimes C_0 + A_0 \otimes B_0 \otimes C_1+ A_0 \otimes B_1 \otimes C_0\\+ A_1 \otimes B_0 \otimes C_0- A_1\otimes B_1 \otimes C_1- A_0 \otimes B_1 \otimes C_1\\- A_1 \otimes B_0 \otimes C_1- A_1 \otimes B_1 \otimes C_0\, , \label{ABC}
 \end{multline} 
in which the six $3\times 3$ Hermitian matrices $A_k$, $B_k$ and $C_k$, \hbox{$k=0,1$},  
satisfy $A^{2}_{k}=B^{2}_{k}=C^{2}_{k}=\mathbb{1}_3$. The Svetlichny inequality~\cite{PhysRevD.35.3066,PhysRevLett.89.060401}
\be 
{\cal I}_3'=\Tr [\eta\, \mathscr{B}'] \leq 4 \label{bell3}
\ee 
tests Bell locality in a tripartite system. The largest possible value in QM is $4 \sqrt{2}$.

We find the six, for simplicity, orthogonal matrices in \eq{ABC} performing a maximization of  ${\cal I}_3'$. These matrices are given in terms of rational numbers as
\bea
A_{0} = -B_1 = -C_0= \left(
\begin{array}{ccc}
 \frac{8}{9} & \frac{4}{9} & \frac{1}{9} \\
 \frac{4}{9} & -\frac{7}{9} & -\frac{4}{9} \\
 \frac{1}{9} & -\frac{4}{9} & \frac{8}{9} \\
\end{array}
\right)\\
A_1=B_0=-C_1=\left(
\begin{array}{ccc}
 \frac{8}{9} & -\frac{4}{9} & \frac{1}{9} \\
 -\frac{4}{9} & -\frac{7}{9} & \frac{4}{9} \\
 \frac{1}{9} & \frac{4}{9} & \frac{8}{9} \\
\end{array}
\right)\, ,
\eea
within a 1\% approximation. We find that $\Gamma=1.372$ yields the maximum value ${\cal I}_3'=4.569$, as computed by using the state in \eq{3state} in the test of \eq{bell3}. 

Experimental determinations of the coefficient $\Gamma$ depend on two Dalitz plot variables 
\be
x= \frac{s_{2}-s_{1}}{m_{{\pi^{+}}}^2} \quad \text{and} \quad y= \frac{s_{3}-s_{0}}{m_{{\pi^{+}}}^2}\, ,
\ee
in which $s_{i}=(p_{j}+p_{k})^{2}$ and $s_{0}=(m_{K_{L}}^{2}+m_{\pi^{+}}^{2}+m_{\pi^{-}}^{2}+m_{\pi^{0}}^{2})/3$. In terms of these, the three-pion decay amplitudes are then written as~\cite{Cirigliano:2011ny}
\begin{multline}
 A(K_L\to \pi^0\pi^0\pi^0) = \\3 (\alpha_1+\alpha_3) + 3 (\zeta_1 - 2\, \zeta_3) \left( y^2 + \frac{1}{3} x^2\right)
 \label{000}
\end{multline}
and
\begin{multline}
 A(K_L\to \pi^0\pi^+\pi^-) =  (\alpha_1+\alpha_3) -(\beta_1+\beta_2) y \\
+ (\zeta_1 - 2\, \zeta_3) \left( y^2 + \frac{1}{3} x^2\right)+(\xi_1 - 2\, \xi_3) \left( y^2 - \frac{1}{3} x^2\right)\, ,
\label{0+-}
\end{multline}
in which the coefficients $\alpha_{i}$, $\zeta_{i}$, $\beta_{i}$ and $\xi_{i}$ are given by chiral perturbation theory. We take their values and related uncertainties from the best fit in~\cite{Bijnens:2002vr}. 
The values of the kaon and pion masses are taken to be $m_{\pi^{\pm}}=139.570$ MeV, $m_{\pi^{0}}=134.977$ MeV and $m_{K^{0}}=497.611$ MeV~\cite{FlaviaNetWorkingGrouponKaonDecays:2010lot}.

\begin{figure}[h!]
\begin{center}
\includegraphics[width=2.5in]{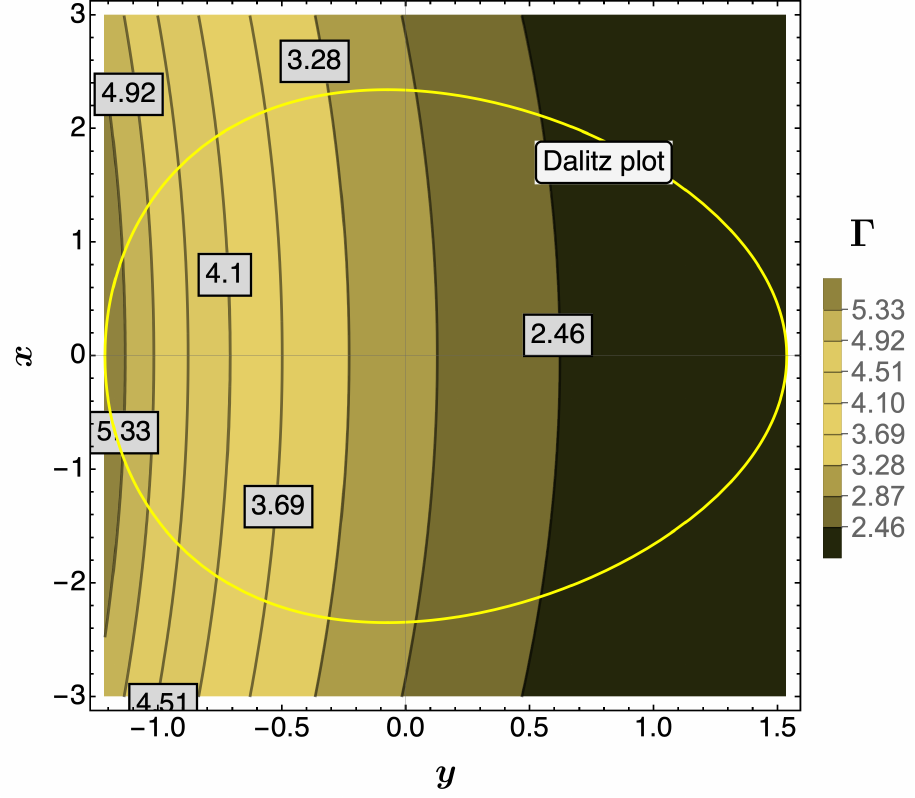}
 \caption{\footnotesize Variation of the coefficient $\Gamma$ of the three-pion state across the kinematic space of the two variables $x$ and $y$ of the  Dalitz plot, which uses the central values~\cite{Bijnens:2002vr} of the coefficients in \eqs{000}{0+-} and refers to the final state containing charged pions---which is the most restrictive. The area inside the closed curve shows the allowed values for the variables $x$ and $y$. 
\label{fig:gamma} }
\end{center}
\end{figure}

\begin{figure}[h!]
\begin{center}
\includegraphics[width=2.5in]{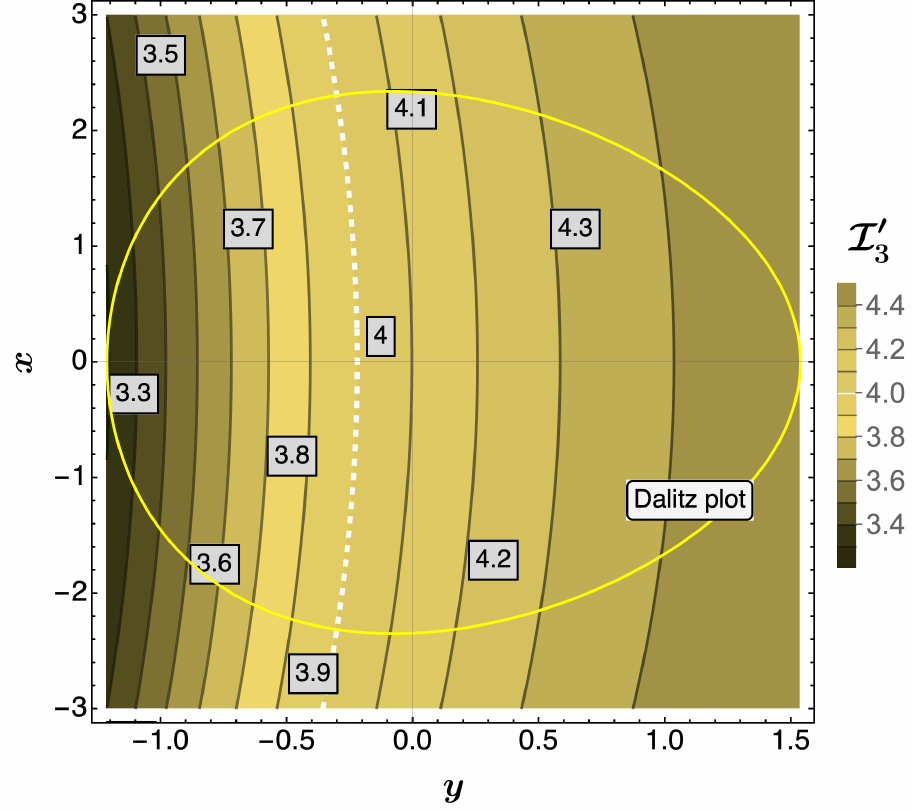}
 \caption{\footnotesize Variation of the value of ${\cal I}_3'$ across the kinematic space of the two variables $x$ and $y$ of the same Dalitz plot as in Fig.~\ref{fig:gamma}. The area inside the closed curve shows the allowed values for the variables $x$ and $y$. The critical value ${\cal I}_3'=4$ is marked by the white dashed contour. \label{fig:i3} }
\end{center}
\end{figure}

Figure~\ref{fig:gamma} depicts the variation of the coefficient $\Gamma$ in \eq{3state} as the variables $x$ and $y$ vary. Only the values within the closed curve are permitted by the kinematics of the decay process.  The value $\Gamma=1.372$ that maximizes the expectation value of Bell operator given in \eq{ABC} lies outside the range of possible values. Figure~\ref{fig:i3} shows the values of ${\cal I}_3'$ as a function of the same Dalitz plot variables. 
For values of the Dalitz plot variable $x$ and $y$ within the region on the left of the white dashed line in 
Fig.~\ref{fig:i3}, the three-pion isospin state does not violate (\ref{bell3}) despite being entangled. This  is possible
 because the system is tripartite, as entanglement inherently implies Bell nonlocality for bipartite pure states. 

The three-pion state in equation \eq{3state} is certified to be entangled and violate non-contextuality and Bell locality inequalities for several values that the Dalitz plot variables within the accessible kinematic region.

Taking as a benchmark the value $\Gamma=2.600 \pm 0.008 $---which is among those allowed in the Dalitz plot and for which  $ {\cal I}_{3}'$ is larger than 4---and whose uncertainty is obtained by propagating those of the coefficients in \eqs{000}{0+-} as given by \cite{Bijnens:2002vr},  we find that the non-contextuality inequality is violated with a very large significance as
\be
 \mathbb{CNTXT}_9 = 3.177\pm 0.001
 \ee
and entanglement is quantified in
  \be
\mathscr{C}[\eta]= 0.979\pm 0.001 \quad \text{and} \quad \quad F_3= 0.959 \pm 0.002
\ee
Moreover,
\be
 {\cal I}_3'=4.257 \pm 0.003
 \ee
 and the  Svetlichny's inequality in \eq{bell3} is also violated with a  significance well above the $5\sigma$ level.

\vskip0.2cm
\textbf{Outlook---}  The isospin space, though more abstract than the ordinary spin space, provides an elegant and simple testing ground for studying the quantum properties of particle states.

The state describing the two pions that emerge from the decay of the neutral kaon $K_{S}$ exhibits quantum contextuality, entanglement, and Bell nonlocality. These quantum properties arise from the interplay between the electro-weak and strong interactions. 

The state describing the decay of the neutral kaon $K_{L}$ into three pions exhibits genuine multipartite entanglement and 
genuine multipartite Bell nonlocality, making it a system of great value for exploring the quantum properties of multipartite states, which are still relatively unexplored.

\vskip0.2cm
\textit{Acknowledgements---} {\small
  LM is supported by the Estonian Research Council under the RVTT3, TK202 and PRG1884 grants.}
\small
\bibliographystyle{JHEP}   
\bibliography{pions.bib} 
\end{document}